\documentclass[a4paper]{jpconf}

\usepackage{graphicx}
\usepackage{iopams}
\usepackage{bm}

\newcommand{\der}[2]{\frac{\partial #1}{\partial #2}}

\newcommand{\be}{\begin{equation}}
\newcommand{\ee}{\end{equation}}
\newcommand{\bea}{\begin{eqnarray}}
\newcommand{\eea}{\end{eqnarray}}
\newcommand{\rtil}{\tilde{r}}
\newcommand{\ttil}{\tilde{t}}
\newcommand{\gm}{\gamma}

\begin{document}

\title{On the local existence of maximal slicings in spherically symmetric 
spacetimes}

\author{Isabel Cordero-Carri\'on$^1$, Jos\'e Mar\'ia Ib\'a\~nez$^1$ and Juan 
Antonio Morales-Lladosa$^1$}

\address{$^1$ Departamento de Astronom\'ia y Astrof\'isica, Universidad de 
Valencia, C/ Dr. Moliner 50, E-46100 Burjassot, Valencia, Spain}

\ead{isabel.cordero@uv.es, jose.m.ibanez@uv.es, antonio.morales@uv.es}

\begin{abstract}
In this talk we show that any spherically symmetric spacetime admits locally a
maximal spacelike slicing. The above condition is reduced to solve a decoupled 
system of first order quasi-linear partial differential equations. The solution 
may be accomplished analytical or numerically. We provide a general procedure 
to construct such maximal slicings.
\end{abstract}

\section{Introduction}

A {\it maximal\,} hypersurface is one that has vanishing mean extrinsic 
curvature, $K = 0$, $K$ being the trace of the extrinsic curvature of the 
hypersurface. The name comes from the fact that the induced volume functional 
reaches a local maximum with respect the variations that keep fixed a given 
boundary. Maximal hypersurfaces were considered by Lichnerowicz \cite{Lichn44} 
to solve Einstein's constraint equations, giving motivation for subsequent 
studies on the subject (see, for example \cite{MmeChoquet76, CaFisMarsdMur76, 
MarsdenTip80, Bartnik84}). In fact, the existence of maximal hypersurfaces is 
extensively used in Mathematical Relativity. This property is a very simple 
geometric assumption to establish general results for broad classes of 
spacetimes, for instance, local or asymptotically stationary or conformally 
flat spacetimes.

We will use the term {\it maximal slicing} when referring to a (non 
intersecting) family of spacelike maximal hypersurfaces which locally foliates 
a certain domain of spacetime. This type of slicing has very nice properties 
as, for example: i) the well-known singularity avoidance capability 
\cite{SmarrY78b}, ii) it is well adapted to the propagation of gravitational 
waves \cite{SmarrY78a, ShibaN95}, and, iii) it gives the natural Newtonian 
analogous when, in addition, conformal flatness is imposed on each slice 
\cite{Isenb78}. Maximal slicing condition has been recently used in the Fully 
Constrained Formulation of Einstein equations derived by the Meudon group 
\cite{BonazGGN04, Lin-Novak06}. 

In spite of their extended use, the existence of maximal slicings in 
spherically symmetric spacetimes (SSSTs) has been only established for vacuum 
and for some particular energy contents (see \cite{Isenb78, Estabrook-WCWST73, 
Beig-Mu98, Reinhart73, Malec94, PetriST85, Eardley-Smarr79}). There is, as far 
as we know, no theorem stating that always it is possible to build a maximal 
slicings in a SSST.

In this work, we aim to prove the local existence of maximal slicings in any 
SSST. We will follow a purely geometrical approach, independent of Einstein 
equations, according to \cite{Reinhart73}, complementary to the standard time 
evolution strategy \cite{Isenb78, Estabrook-WCWST73, Beig-Mu98, Malec94, 
PetriST85, Eardley-Smarr79}. 

Although our study is independent of the field equations, one by-product of our
approach, which could be of interest in the field of Numerical Relativity, is 
that it provides a means to assess complex and sophisticated 3D numerical codes
built to solve Einstein equations.

\section{Local existence of maximal slicings}

In this section we establish the following result:

\noindent {\bf Theorem.} {\it Any spherically symmetric spacetime can be 
locally sliced by a family of maximal spacelike hypersurfaces.}

In order to prove the previous theorem, we derive a decoupled system of three 
first order partial differential equations that proves the local existence of a 
maximal slicings, and provides a general procedure allowing its construction. 

Let us start with the canonical form of the metric of a SSST,
\be
    ds^2 = A\,dt^2 + 2C\,dt\ dr + B\,dr^2 + D\,d\Omega^2,
\label{e:sph_sym}
\ee
where $d\Omega^2=d\theta^2 + \sin^2\theta d\varphi^2$ is the metric of the 
2-sphere, $A, B, C, D$ are smooth functions of $t$ and $r$, and $AB-C^2<0$ to 
ensure the Lorentzian character of the metric. In addition we choose the 
signature $(-,+,+,+)$, and accordingly $D>0$. Partial derivatives with respect 
to $r$ will be denoted as $\displaystyle{\der{f}{r} = f'}$, and with respect to
$t$ as $\displaystyle{\der{f}{t} = \dot{f}}$. The spatial metric $\gm_{ij}$ 
induced on the hypersurfaces $\Sigma_t$, defined by $t=\mathrm{constant}$, is 
$\gm_{ij} = \mathrm{diag}(B, D,D\sin^2\theta)$, where $B>0$ since we are 
considering spacelike hypersurfaces. Let $n$ be the future pointing timelike 
unit normal to the hypersurfaces $\Sigma_t$,
\be
    n = \frac{1}{\alpha}\left(\der{}{t} - \frac{C}{B}\der{}{r}\right), \;\; 
\alpha = \sqrt{\frac{C^2}{B}-A}.
\label{e:def_n}
\ee
The mean extrinsic curvature $K$ of $\Sigma_t$ is related with the expansion of
$n$, $K = -\nabla \cdot n$, where $\nabla$ is the covariant derivative with 
respect to the spacetime metric. In the given metric in Eq.~(\ref{e:sph_sym}), 
this relation is
\be
    K = \frac{1}{2\alpha B} \left( - \dot{B} - 2B\frac{\dot{D}}{D} + 2C' 
- C\frac{B'}{B} + 2C\frac{D'}{D} \right).
\label{e:K}
\ee

In the following, we assume that $A, B, C$ and $D$ are known functions. We look
for a change of coordinates 
$\left\{\ttil=\ttil(t,r),\rtil=\rtil(t,r),\theta,\varphi\right\}$ such that the
hypersurfaces $\ttil=\mathrm{constant}$ are maximal. We introduce two fields, 
$X$ and $Y$, satisfying the commutation relation $[X, Y]=0$. This condition 
assures the existence of two coordinate parameters, namely $\ttil=\ttil(t,r)$ 
and $\rtil=\rtil(t,r)$, such that
\be
    X = \der{}{\ttil},\; Y = \der{}{\rtil}.
\label{e:def_X_Y}
\ee
Then, we decompose these fields as $\displaystyle Y = \lambda \bar{Y}, \,\,
X = a \bar{Y} + b \bar{Y}^{\bot}$, with $\bar{Y}^2=1$, 
$\bar{Y}\cdot\bar{Y}^{\bot}=0$, $b\neq0$, and $\lambda>0$. The condition 
$\bar{Y}^2=1$ is equivalent to
\be
    \bar{Y} = f \der{}{t} + P \der{}{r}, \;\;P = B^{-1}\left( - f C
+ \epsilon \sqrt{f^2 l^2 + B} \right),
\label{e:def_P}
\ee
being $f$ an unknown function to be determined, $\epsilon=\pm1$ and 
$l^2=-AB+C^2>0$. Fixing the coefficient of $\displaystyle \der{}{t}$ in the 
decomposition of $\bar{Y}^{\bot}$, $\bar{Y}\cdot\bar{Y}^{\bot}=0$ leads to
\be
    \bar{Y}^{\bot} = \der{}{t} + Q \der{}{r},\;Q = B^{-1}\left( - C
+ \frac{\epsilon f l^2}{\sqrt{f^2 l^2 + B}} \right).
\label{e:def_Q}
\ee
Consequently, the resulting fields are
\be
    X = (af+b)\der{}{t} + (aP+bQ)\der{}{r}, \,\,\, 
    Y = \lambda\left(f\der{}{t} + P\der{}{r}\right) = \lambda\left(f\alpha\,n
+ \frac{\epsilon}{B}\sqrt{f^2l^2+B}\,\der{}{r} \right), \label{e:dec_X_Y}
\ee
where we have taken into account Eqs.~(\ref{e:def_n}) and (\ref{e:def_P}). The 
condition $[X,Y]=0$ is then equivalent to
\be
	\left[\frac{P}{b\,p}\right]' = - \left[\frac{f}{b\,p}\right] 
\mbox{\LARGE $^{^.}$}, \,\,\,\,
- \left[\frac{a P + b\,Q}{b\,p\,\lambda}\right]' = 
\left[\frac{a f + b}{b\,p\,\lambda}\right] \mbox{\LARGE $^{^.}$},
\label{e:conm_rel}
\ee
where $\displaystyle p = P-fQ = \frac{\epsilon}{\sqrt{f^2l^2+B}} \neq 0$.

Now, we denote with $\widetilde{K}$ the trace of the extrinsic curvature of the 
new hypersurfaces $\ttil=\mathrm{constant}$. The condition $\widetilde{K} = 0$ 
and the commutation relation provide 3 equations for 4 unknown functions, 
$a,b,f,\lambda$. Taking into account that $D$ is a scalar under the above 
change of coordinates, we can add, without loss of generality, the following 
coordinate condition
\be
    \rtil^2\,Y^2 = D,
\label{e:def_conf_flat}
\ee
saying that the metric on the hypersurfaces $\ttil=\mathrm{constant}$ is 
written in isotropic conformally flat form. From Eqs.~(\ref{e:K}) and 
(\ref{e:def_conf_flat}), the condition $\widetilde{K} = 0$ is equivalent to
\be
    2 Y \left( X \cdot Y \right) - 3 X \left(Y^2\right) + \left[\frac{4}{\rtil}
+ \frac{Y\left(Y^2\right)}{Y^2} \right] X\cdot Y = 0.
\label{e:max_sli}
\ee

From the decompositions (\ref{e:dec_X_Y}), Eqs.~(\ref{e:def_conf_flat}) and 
(\ref{e:max_sli}) are expressed as $\lambda = \sqrt{D} / \rtil$, that can be 
viewed as a definition of $\rtil$ in terms of $\lambda$, and as
\be
    f \dot{a} + P a' - a \left( \frac{f \dot{\lambda} + P \lambda'}{\lambda}
-\frac{2}{\sqrt{D}} \right) = 3 b \;\frac{\dot{\lambda} + Q \lambda'}{\lambda}.
\label{e:der_a_lambda}
\ee

After some algebraic calculations, the previous definition of $\lambda$ and 
Eqs.~(\ref{e:conm_rel}) and (\ref{e:der_a_lambda}) are equivalent to
\be
	a = b \sqrt{D} \left(\frac{\dot{\lambda}+Q\lambda'}{\lambda}
- \frac{\dot{D}+QD'}{2D} \right),
\label{e:a}
\ee
\be
	f \frac{\dot{\lambda}}{\lambda} +  P \frac{\lambda'}{\lambda} = 
f \frac{\dot{D}}{2D} + P \frac{D\,'}{2D} - \frac{1}{\sqrt{D}} \, ,
\label{e:lambda}
\ee
\be
	f \frac{\dot{b}}{b} + P \frac{b\,'}{b} = P' - P \frac{p\,'}{p} + \dot{f}
- f \frac{\dot{p}}{p} \, ,
\label{e:b}
\ee
and
\be
	\frac{\dot{p}}{p} - \frac{\dot{D}}{D} - Q' + Q \left[ \frac{p\,'}{p} 
- \frac{D'}{D} \right] = 0.
\label{e:f}
\ee
Notice that Eq.~(\ref{e:f}) involves only $f$ when $p$ and $Q$ are written 
explicitly in terms of $f$.

First, Eq.~(\ref{e:f}) can be solved for $f$. Second, Eqs.~(\ref{e:lambda}) and
(\ref{e:b}) can be solved for $\lambda$ and $b$. Finally, $a$ can be obtained 
from Eq.~(\ref{e:a}). Assuming that $A, B, C, D$ are continuously 
differentiable functions, the initial value problem with respect to this set of
equations has always local (both in space and time) solution \cite{pde} (which 
is also continuously differentiable). Therefore, we have proved the announced 
theorem.

Notice that in order to solve this set of equations, it can be useful to 
distinguish two different cases, $f=0$ and $f\neq0$. In the case of $f=0$, 
Eq.~(\ref{e:f}) is reduced to $K=0$, and the rest of equations can be 
integrated easily. In the case of $f\neq0$, it can be defined the variable 
$\displaystyle F = \frac{\epsilon f}{\sqrt{f^2l^2+B}} \Leftrightarrow 
f = \epsilon F\sqrt{\frac{B}{1-l^2F^2}}$, and Eqs.~(\ref{e:lambda}), 
(\ref{e:b}) and (\ref{e:f}) can be rewritten as a hyperbolic system of 
equations for $F$, $\lambda$ and $b$. 

\section{Conclusions}
Two basic results have been displayed: i) A theorem ensuring the existence of 
maximal slicings in any SSST. ii) A geometrical method to build up such slices 
by solving three decoupled first order quasi-linear partial differential 
equations (\ref{e:lambda}), (\ref{e:b}) and (\ref{e:f}). The first result aims 
to fill a theoretical gap in the scientific literature. The second one tries to
achieve an algorithmic procedure to obtain maximal slicings. An interesting 
by-product for Numerical Relativity of the approach presented in this paper has
to do with the assessment of 3D codes written, as customary, in Cartesian 
coordinates. Let us consider two codes NC1 and NC2 such that only NC1 uses a 
gauge which is maximal. Hence, the evolution with code NC2 of any initial data 
admitting a spherically symmetric limit could be compared to the evolution 
produced by code NC1, by simply using our procedure to generate a SSST 
satisfying the maximal slicing condition.

\ack
I. C.-C. acknowledges support from the Spanish Ministerio de Educaci\'on y 
Ciencia (MEC) (AP2005-2857). This work has been also supported by the MEC Grant
No. AYA2007-67626-C03-01, the MEC-FEDER project No. FIS2006-06062 and the 
MICIN-FEDER project No. FIS2009-07705.


\section*{References}

\begin{thebibliography}{9}
%
\bibitem{Lichn44}
Lichnerowicz A (1944) {\it J. Math. Pures Appl.} {\bf 23} 37
%
\bibitem{MmeChoquet76}
Choquet-Bruhat Y (1976) {\it Annali della Scuola Normale Superiore di Pisa} 
{\em Classe di Scienze $4^e$ série}, tome 3, nº3 361
%
\bibitem{CaFisMarsdMur76}
Cantor M, Fisher A, Marsden J, Murchadha N $\bar{\rm{O}}$ and York J (1976) 
{\it Commun. Math. Phys.} {\bf 49} 187
%
\bibitem{MarsdenTip80}
Marsden J E and Tipler F J (1980) {\it Phys. Rep.} {\bf 66} 109
%
\bibitem{Bartnik84}
Bartnik R (1984) {\it Commun. Math. Phys.} {\bf 94} 155
%
\bibitem{SmarrY78b}
Smarr L and  York J W (1978) {\it Phys. Rev. D} {\bf 17} 2529
%
\bibitem{SmarrY78a}
Smarr L and  York J W (1978) {\it Phys. Rev. D} {\bf 17} 1945
%
\bibitem{ShibaN95}
Shibata M and Nakamura T (1995) {\it Phys. Rev. D} {\bf 52} 5428
%
\bibitem{Isenb78}
Isenberg J A (2008) {\it Int. J. Mod. Phys. D} {\bf 17} 265
%
\bibitem{BonazGGN04}
Bonazzola S, Gourgoulhon E, Grandcl\'ement P and Novak J (2004) 
{\it Phys. Rev. D} {\bf 70} 104007
%
\bibitem{Lin-Novak06}
Lin L-M and Novak J (2006) {\it Class. Quantum Grav.} {\bf 23} 4545
%
\bibitem{Estabrook-WCWST73}
Estabrook F, Wahlquist H, Christensen S, DeWitt B, Smarr L and Tsiang E (1973) 
{\it Phys. Rev. D} {\bf 7} 2814
%
\bibitem{Beig-Mu98}
Beig R and Murchadha N \'{O} (1998) {\it Phys. Rev. D} {\bf 57} 4728
%
\bibitem{Reinhart73}
Reinhart B L (1973) {\it J. Math. Phys.} {\bf 14} 719
%
\bibitem{Malec94}
Malec E (1994) {\it Phys. Rev. D} {\bf 49}, 6475
%
\bibitem{PetriST85}
Petrich L I, Shapiro S L and Teukolsky S A (1985) {\it Phys. Rev. D} {\bf 31}, 
2459
%
\bibitem{Eardley-Smarr79}
Eardley D M  and Smarr L (1979) {\it Phys. Rev. D} {\bf 19} 2239
%
\bibitem{pde}
Zauderer E (1989) {\it Partial differential equations of applied mathematics}, 
Second edition (John Wiley \& Sons, New York)
%
\end{thebibliography}
\end{document}